\documentclass[runningheads]{llncs}
\newcommand{\Figure}[1]{Figure~\ref{fig:#1}}

\newcommand{\Table}[1]{Table~\ref{tab:#1}}

\newcommand{\Eq}[1]{Eq.~\ref{eq:#1}}

\newcommand{\Section}[1]{Section~\ref{sec:#1}}

\usepackage{amsmath}
\usepackage{bbold}
\renewcommand\vec{\mathbf}
\newcommand{\mat}{\mathbf}

\DeclareMathOperator*{\argmin}{argmin}

\usepackage{enumitem}

\usepackage{caption}
\usepackage{subcaption}

\usepackage{multirow}
\usepackage{graphicx}

\usepackage{sidecap}

\usepackage{float}

\usepackage[flushleft]{threeparttable}

\usepackage{amssymb}

\usepackage[linesnumbered,ruled,vlined]{algorithm2e}
\usepackage {algpseudocode}
\usepackage{algorithmicx}
\usepackage{algcompatible}

\newcommand{\nosemic}{\renewcommand{\@endalgocfline}{\relax}}
\newcommand{\dosemic}{\renewcommand{\@endalgocfline}{\algocf@endline}}
\let\oldnl\nl
\newcommand{\nonl}{\renewcommand{\nl}{\let\nl\oldnl}}
\usepackage{hyperref}
\hypersetup{
    colorlinks=true,
    linkcolor=red,
    filecolor=magenta,      
    urlcolor=blue,
    pdftitle={Overleaf Example},
    pdfpagemode=FullScreen,
    }

\usepackage{graphicx}
\begin{document}
\title{Deep Cardiac MRI Reconstruction with ADMM}
%
%
\author{George Yiasemis \inst{1,2},
Nikita Moriakov\inst{1,2},
Jan-Jakob Sonke\inst{1,2},
Jonas Teuwen\inst{1,2,3}\\
\email{\{g.yiasemis, n.moriakov, j.sonke, j.teuwen\}@nki.nl}  }

\authorrunning{G. Yiasemis et al.}
%
\institute{Netherlands Cancer Institute, Amsterdam, Netherlands \\
\and
University of Amsterdam,  Amsterdam, Netherlands\\
\and
Radboud University Medical Center, Nijmegen, Netherlands\\
}


\maketitle              
\begin{abstract}
Cardiac magnetic resonance imaging (CMR) is a valuable non-invasive tool for identifying cardiovascular diseases. For instance, Cine MRI is the benchmark modality for assessing the cardiac function and anatomy. On the other hand, multi-contrast (T1 and T2) mapping has the potential to assess pathologies and abnormalities in the myocardium and interstitium. However, voluntary breath-holding and often arrhythmia, in combination with MRI's slow imaging speed, can lead to motion artifacts, hindering real-time acquisition image quality. Although performing accelerated acquisitions can facilitate dynamic imaging, it induces aliasing, causing low reconstructed image quality in Cine MRI and inaccurate T1 and T2 mapping estimation. In this work, inspired by related work in accelerated MRI reconstruction, we present a deep learning-based method for  accelerated cine and multi-contrast reconstruction in the context of dynamic cardiac imaging. We formulate the reconstruction problem as a least squares regularized optimization task, and employ vSHARP, a state-of-the-art Deep Learning-based inverse problem solver, which incorporates half-quadratic variable splitting and the alternating direction method of multipliers (ADMM) with neural networks. We treat the problem in two setups; a 2D reconstruction and a 2D dynamic reconstruction task, and employ 2D and 3D deep learning networks, respectively. Our method optimizes in both the image and k-space domains, allowing for high reconstruction fidelity. Although the target data is undersampled with a Cartesian equispaced scheme, we train our deep neural network using both Cartesian and simulated non-Cartesian undersampling schemes to enhance generalization of the model to unseen data, a key ingredient of our method. Furthermore, our model adopts a deep neural network to learn and refine the sensitivity maps of multi-coil k-space data. Lastly, our method is jointly trained on both, undersampled cine and multi-contrast data.

\end{abstract}
\newpage
\section{Introduction}
\label{sec:sec1}

Cardiac magnetic resonance (CMR) stands as a vital clinical tool for assessing cardiovascular diseases due to its non-invasive and radiation-free nature, enabling a comprehensive evaluation of cardiovascular aspects, such as structure, function, flow, perfusion, viability, tissue characterization, as well as the assessment of myocardial fibrosis and other pathologies \cite{Arai2011,Kim2017,Larose2007}. Key CMR applications  include cine MR imaging and T1/T2 mapping. 

However,  CMR faces inherent physical challenges, primarily the time consuming MRI acquisition process.  The requirement for increased spatiotemporal resolution in cardiac imaging further amplifies this challenge. To mitigate prolonged scan times, accelerated MRI acquisitions are utilized by obtaining undersampled $k$-space data, though this approach violates the Nyquist-Shannon sampling criterion  \cite{1697831}.

In the broader MRI domain, conventional techniques such as  Parallel Imaging (PI) \cite{Griswold2002,Niendorf2006} and Compressed Sensing (CS) \cite{Geethanath2013,Kido2016} have been employed to accelerate MRI data acquisition. These approaches leverage spatial sensitivity information from multiple receiver coil arrays and exploit the sparsity or compressibility of MRI data.  However, these methods have limitations, such as noise amplification in PI, and assumptions of sparsity that may not hold for all MRI data in CS, whilst finding optimal parameters for CS methods might be computationally and time consuming.

In the last decade, Deep Learning (DL) has revolutionized MRI image reconstruction, exhibiting superior performance compared to traditional methods, especially in accelerated MRI reconstruction tasks \cite{pal2022review}.  DL-based algorithms can learn complex image representations directly from available datasets, enabling enhanced image reconstruction from undersampled $k$-space measurements, often in supervised learning \cite{10.3389/fnins.2022.919186,Kstner2020,10.1007/978-3-030-59713-9_7,Yiasemis_2022_CVPR}, or self-supervised settings \cite{Hamilton2022}. This advancement holds significant potential to impact CMR by elevating the image quality of reconstructed highly undersampled data while concurrently reducing breath-hold duration.

In this work, motivated by the need for reducing acquisition times and breath-hold durations further during CMR, we employ vSHARP \cite{yiasemis2023vsharp} (variable Splitting Half-quadratic ADMM algorithm for Reconstruction of inverse-Problems), a DL-based inverse problem solver, previously applied on brain and prostate MR imaging exhibiting state-of-the-art performance. We particularize vSHARP for accelerated Cardiac MRI Reconstruction and introduce in \Section{subsubsec3.1.2} two variants by treating the problem at hand as a 2D reconstruction task (2D model) or as a 2D dynamic reconstruction task (3D model). Additionally, in \Section{subsec3.2} we propose various training techniques to boost model training and generalizability across unseen cardiac (cine and T1/T2) MRI data. In \Section{sec5}, we experimentally compare our two approaches, highlighting that our 2D dynamic implementation outperforms traditional 2D reconstruction and we further compare our models with current state-of-the-art approaches.

\section{Theory and Problem Formulation}
\subsection{Accelerated MRI Reconstruction}
\label{sec:subsec2.1}
Recovering a two-dimensional image $\Vec{x}^{*}\in\mathbb{C}^{N}$ from undersampled  multi-coil (assume $N_c$ coils) $k$-space measurements $\tilde{\Vec{y}}\in \mathbb{C}^{N\times N_{c}}$ can be formulated as a minimisation problem as follows:
\begin{equation}
    \Vec{x}^{*} = \argmin_{\vec{x}\in\mathbb{C}^{N}}\frac{1}{2} \sum_{k=1}^{N_c}\left|\left| \mathcal{A}^{k}(\vec{x}) - \Tilde{\vec{y}}^{k}\right|\right|_2^2 + \mathcal{R}(\vec{x}), \quad \mathcal{A}^{k} =  \mat{U} \mathcal{F} \mat{S}^{k},
\label{eq:variational_problem}
\end{equation}
where $\mathcal{A}^{k}$ represents the forward or corruption operator per coil. It involves mapping the image to an individual coil image using a known sensitivity map $\mat{S}^{k}$, transforming it to the $k$-space domain via the Fast Fourier Transform (FFT) $\mathcal{F}$, and undersampling with $\mat{U}$. The function $\mathcal{R}: \mathbb{C}^{N} \to \mathbb R$ denotes a regularization functional, which is assumed to impose prior knowledge about the image.

In the context of cardiac magnetic resonance, acquisitions are typically dynamic and synchronized with electrocardiography (ECG)-derived cardiac cine. In dynamic acquisitions, multiple undersampled $k$-space data $\tilde{\Vec{y}}\in \mathbb{C}^{N\times N_{c} \times N_{f}}$ are obtained at $N_{f}$ time frames. Consequently, \Eq{variational_problem} is adapted as follows:
\begin{equation}
    \Vec{x}^{*}_{\text{d}} = \argmin_{\vec{x}\in\mathbb{C}^{N \times N_{f}}}\frac{1}{2} \sum_{t=1}^{N_f}\sum_{k=1}^{N_c}\left|\left| \mathcal{A}^{k}(\vec{x}_{\cdot, t}) - \Tilde{\vec{y}}_{\cdot, t}^{k}\right|\right|_2^2 + \mathcal{R}(\vec{x}), \quad \mathcal{A}^{k} =  \mat{U} \mathcal{F} \mat{S}^{k}.
\label{eq:variational_problem_dynamic}
\end{equation}
In dynamic acquisitions, it is often assumed that knowledge can be shared across time frames or that the motion pattern is known, thereby requiring the selection of an appropriate prior $\mathcal{R}: \mathbb{C}^{N \times N_{f} } \to \mathbb R$ that incorporates this information \cite{Ye2019}.

\section{Methods}
\label{sec:sec3}

\subsection{Deep Learning Framework}
\label{sec:subsec3.1}

\subsubsection{Sensitivity Map Prediction}
\label{sec:subsubsec3.1.1}
In conventional settings, sensitivity maps are estimated from the autocalibration signal (ACS) data, often incorporating a portion of the center of the $k$-space. Advanced techniques for refining these estimated sensitivities include ESPIRiT or GRAPPA \cite{Griswold2002,Uecker2013}. However, these approaches can impose computational constraints. To overcome the need for such computationally expensive algorithms, we employ a two-dimensional deep learning module, specifically a 2D U-Net \cite{Ronneberger2015}. This model takes ACS-estimated sensitivity maps as input and produces refined versions of them as output. The predicted sensitivity maps $\left\{ \mat{S}^{k} \right\}_{k=1}^{N_c}$ are used for downstream reconstruction tasks, and the sensitivity module is trained in an end-to-end manner along with the reconstruction model.

\subsubsection{Reconstruction via ADMM Unrolled Optimization}
\label{sec:subsubsec3.1.2}
Our approach utilizes vSHARP \cite{yiasemis2023vsharp}, a DL-based inverse problem solver, to address \Eq{variational_problem}. vSHARP employs the half-quadratic variable splitting method \cite{Li2020} to transform the optimization problem in \Eq{variational_problem} by introducing an intermediate variable $\vec{w}$. It then unrolls the optimization process over $T$ iterations using the alternating direction method of multipliers algorithm (ADMM) \cite{boyd2011distributed}, as follows:

\begin{subequations}
\begin{equation}
    \vec{w}^{(j+1)} = \argmin_{\vec{w}\in\mathbb{C}^{N}} \mathcal{R}(\vec{w}) + \frac{\lambda}{2} \big | \big | \vec{x}^{(j)} - \vec{w} + \frac{\vec{m}^{(j)}}{\lambda} \big | \big |_2^2,
\label{eq:admm_w}
\end{equation}
\begin{equation}
    \vec{x}^{(j+1)} = \argmin_{\vec{x}\in\mathbb{C}^{N}} \frac{1}{2} \sum_{k=1}^{N_c}\left|\left| \mathcal{A}^{k}(\vec{x}) - \Tilde{\vec{y}}^{k}\right|\right|_2^2 + \frac{\lambda}{2} \big | \big | \vec{x} - \vec{w}^{(j+1)} + \frac{\vec{m}^{(j)}}{\lambda} \big | \big |_2^2,
\label{eq:admm_x}
\end{equation}
\begin{equation}
    \vec{m}^{(j+1)} = \vec{m}^{(j)} + \lambda (\vec{x}^{(j+1)} - \vec{w}^{(j+1)}), \quad j=0,\cdots, T-1.
\label{eq:admm_m}
\end{equation}
\label{eq:admm}
\end{subequations}

Our method incorporates U-Nets to replace the need for manually selecting a prior functional $\mathcal{R}$ in \Eq{admm_w} and learn the solution from data directly, namely the denoising step. Next, data consistency is enforced by solving \Eq{admm_x}  via an unrolled (differentiable) gradient descent scheme. Our approach initializes $\vec{w}^{(0)}$ and $\vec{x}^{(0)}$ using a zero-filled reconstruction with $\Tilde{\vec{y}}$ and the predicted coil sensitivity maps: $\vec{w}^{(0)} = \vec{x}^{(0)} := \sum_{k=1}^{N_c}{\mat{S}^{k}}^{*} \mathcal{F}^{-1} (\Tilde{\mat{y}}^{k}) $. Additionally, a learned initializer, adapted from \cite{Yiasemis_2022_CVPR}, is used to determine an initialization for the Lagrange Multipliers $\vec{m}^{(0)}$. For dynamic reconstruction as in \Eq{variational_problem_dynamic}, \Eq{admm} is replaced by:
\begin{subequations}
\begin{equation}
    \vec{w}^{(j+1)} = \argmin_{\vec{w}\in\mathbb{C}^{N\times N_f}} \mathcal{R}(\vec{w}) + \frac{\lambda}{2} \big | \big | \vec{x}^{(j)} - \vec{w} + \frac{\vec{m}^{(j)}}{\lambda} \big | \big |_2^2,
\label{eq:admm_w_3d}
\end{equation}
\begin{equation}
    \vec{x}^{(j+1)} = \argmin_{\vec{x}\in\mathbb{C}^{N\times N_f}} \frac{1}{2} \sum_{t=1}^{N_f}\sum_{k=1}^{N_c}\left|\left| \mathcal{A}^{k}(\vec{x}_{\cdot, t}) - \Tilde{\vec{y}}_{\cdot, t}^{k}\right|\right|_2^2 + \frac{\lambda}{2} \big | \big | \vec{x} - \vec{w}^{(j+1)} + \frac{\vec{m}^{(j)}}{\lambda} \big | \big |_2^2,
\label{eq:admm_x_3d}
\end{equation}
\begin{equation}
    \vec{m}^{(j+1)} = \vec{m}^{(j)} + \lambda (\vec{x}^{(j+1)} - \vec{w}^{(j+1)}), \quad j=0,\cdots, T-1.
\label{eq:admm_m_3d}
\end{equation}
\label{eq:admm_3d}
\end{subequations}

\subsection{Model Training Techniques}
\label{sec:subsec3.2}
In this section, we outline the various additional techniques employed in our paper to enhance the performance of our models.

\subsubsection{Joint Modality Training}
\label{sec:subsubsec3.2.1}
During the training of our DL-based approach, we jointly trained it using all available data at our disposal (see \Section{subsec4.3}). This approach served a dual purpose; Firstly, instead of training separate models for each modality, our joint modality training aimed to utilize a larger dataset promoting more effective learning and generalization. Moreover, by integrating cine and T1/T2-weighted MRI data, we aimed to harness the  complementarity between these modalities. This approach enabled the model to exploit the shared features and correlations, potentially improving the reconstruction quality for both modalities.

\subsubsection{Random $k$-space Cropping}
\label{sec:subsubsec3.2.2}

\begin{figure}[!htb]
\centering
\includegraphics[width=0.9\textwidth]{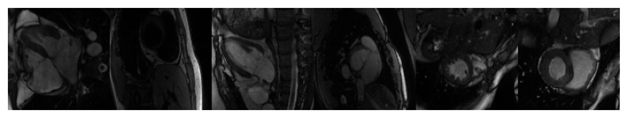}
\caption{Randomly cropped (in the image domain) examples of cine and T1/T2-weighted MRI images from the dataset. These images are then transformed to the $k$-space domain, followed by retrospective undersampling, and are subsequently utilized for training.}
\label{fig:crops}
\end{figure}

To optimize computational efficiency during training, we utilized random cropping on the fully-sampled multi-coil $k$-space data. Since direct cropping of the $k$-space would be inappropriate, we first applied the inverse Fast Fourier Transform (FFT) to reconstruct it into fully-sampled multi-coil images. Subsequently, random cropping was performed on this reconstructed image, and the resulting cropped image was transformed back to the $k$-space domain (via FFT). The $k$-space data was then undersampled and used as input to our model. This approach not only offered computational benefits but also allowed our model to gain exposure to different parts of the reconstructed data, including background noise and the regions of interest, without compromising overall reconstruction quality as compared to using non-cropped data. \Figure{crops} illustrates examples of cropped images before the transformation back to the $k$-space domain. It's important to note that for dynamic data, the same cropping process was applied to all time frames.

\subsubsection{Multi-Scheme Undersampling}
\label{sec:subsubsec3.2.3}
\begin{figure}[!htb]
\centering
\includegraphics[width=0.9\textwidth]{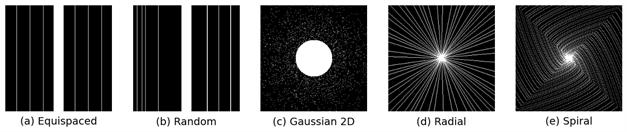}
\caption{Undersampling Schemes during training.}
\label{fig:schemes}
\end{figure}

Undersampling for the target (validation) data comprised Cartesian rectilinear equispaced undersampling masks, with 24 fully-sampled ACS (central) lines, and with acceleration factors of $R=$ 4, 8 and 10. Inspired by previous work \cite{yiasemis2023retrospective},  which demonstrated enhanced model generalizability in reconstructing Cartesian rectilinear data, we employed a multi-scheme undersampling setup during training. Alongside the provided undersampling pattern, we used the following undersampling schemes: Equispaced and Random Cartesian rectilinear, Gaussian 2D Cartesian, and pseudo-Radial and pseudo-Spiral schemes. These undersampling schemes are visualized in \Figure{schemes}. Note that for dynamic data, the same undersampling scheme was applied on all time frames.

\subsubsection{Dual Domain Loss}
\label{sec:subsubsec3.2.4}
To train our models we designed a dual-domain loss:
\begin{equation}
    \mathcal{L}_{\boldsymbol{\phi}} = \mathcal{L}_{\boldsymbol{\phi}}^{img} + \mathcal{L}_{\boldsymbol{\phi}}^{freq},
\end{equation}
where $\mathcal{L}_{\boldsymbol{\phi}}^{img}$ and $ \mathcal{L}_{\boldsymbol{\phi}}^{freq}$ represent losses computed in the image and frequency domain, respectively.

\paragraph{Image Domain Loss}
\label{sec:para3.2.4.1}

The image domain loss, $\mathcal{L}_{\boldsymbol{\phi}}^{img}$, is computed between the ground truth RSS image $\vec{x}$ and the magnitude of the model-predicted image $\hat{\vec{x}}_{\boldsymbol{\phi}}$. This loss comprises several components:
\begin{equation}
    \mathcal{L}_{\boldsymbol{\phi}}^{img} = \lambda_{
    \text{SSIM}}\mathcal{L}_\text{SSIM}\left(\vec{x},\, \hat{\vec{x}}_{\boldsymbol{\phi}}\right) +  \lambda_{1} \mathcal{L}_{1} \left(\vec{x},\, \hat{\vec{x}}_{\boldsymbol{\phi}}\right) \\
    + \lambda_{\text{HFEN}_{1}}\mathcal{L}_{\text{HFEN}_{1}} \left(\vec{x},\, \hat{\vec{x}}_{\boldsymbol{\phi}}\right) 
    \label{eq:img_loss}
\end{equation}
which are defined as follows:
\begin{equation}
\begin{gathered}
        \mathcal{L}_{\text{SSIM}} (\vec{u},\,\vec{v}) = 1- {\text{SSIM}} (\vec{u},\,\vec{v}), \quad \mathcal{L}_{1}  (\vec{u},\,\vec{v}) = \left|\left| \vec{u} - \vec{v} \right| \right|_1, \\
        \text{and,} \quad  \mathcal{L}_{\text{HFEN}_1}(\vec{u},\, \vec{v})
        =  \text{{HFEN}}_1(\vec{u},\, \vec{v}).
    \label{eq:img_losses}
\end{gathered}
\end{equation}

In \Eq{img_losses}, SSIM denotes the Structural Similarity Index Measure, computed over W windows, each of size $7\times 7$ pixels extracted from images $\vec{u}$ and $\vec{v}$. It is defined as:
\begin{equation}
    \text{SSIM}(\vec{u},\,\vec{v}) =
    \frac{1}{W}\sum_{i=1}^{W} \frac{(2\mu_{\vec{u}_i}\mu_{\vec{v}_i} + 0.01)(2\sigma_{\vec{u}_i\vec{v}_i} + 0.03)}{({\mu^2_{\vec{u}_i}} +{\mu^2_{\vec{v}_i}} + 0.01)({\sigma^2_{\vec{u}_i}} + {\sigma^2_{\vec{v}_i}} + 0.03)}.
    \label{eq:ssim_metric}
\end{equation}
Here, $\mu_{\vec{u}_i}$, $\mu_{\vec{v}_i}$, $\sigma_{\vec{u}_i}$ and $\sigma_{\vec{v}_i}$ represent the means and standard deviations of each window, while $\sigma_{\vec{u}_i\vec{v}_i}$ signified the covariance between $\vec{u}_i$ and $\vec{v}_i$. HFEN$_1$ represents the High-Frequency Error Norm, and is defined as follows:
\begin{equation}
    \text{{HFEN}}_1(\vec{u},\, \vec{v})\, = \, \frac{|| \text{{G}}(\vec{u}) - \text{{G}}(\vec{v}) ||_1}{||\text{{G}}(\vec{u})||_1},
    \label{eq:hfen}
\end{equation}
where $\text{G}$ denotes a $15\times 15$ Laplacian of Gaussian filter with a standard deviation of 2.5.

SSIM and HFEN are computed per single 2D slice/time frame.  For dynamic reconstruction experiments, we also incorporated $\lambda_{\text{SSIM3D}}  \mathcal{L}_{\text{SSIM3D}}$, which computes the SSIM metric for volumes using windows of voxel-size $7\times 7 \times 7$. 

\paragraph{Frequency Domain Loss}
\label{sec:para3.2.4.2}
The frequency domain loss, $\mathcal{L}_{\boldsymbol{\phi}}^{freq}$, was computed between the ground truth multi-coil $k$-space $\vec{y}$ and the $k$-space transformation of the model predicted image $\hat{\vec{y}}_{\boldsymbol{\phi}}$:
\begin{equation}
    \mathcal{L}_{\boldsymbol{\phi}}^{freq} = \lambda_{\text{NMAE}}\mathcal{L}_{\text{NMAE}}\left(\vec{y},\, \hat{\vec{y}}_{\boldsymbol{\phi}}\right), \text{ where } \mathcal{L}_\text{NMAE} (\vec{u},\, \vec{v})\,= \, \frac{||\vec{u}\,-\,\vec{v}||_1}{||\vec{u}||_1}.
    \label{eq:freq_loss}
\end{equation}
The choice of  the weighting factors $\lambda_{\text{SSIM}}$, $\lambda_{\text{SSIM3D}}$, $\lambda_{1}$, $\lambda_{\text{HFEN}_{1}}$, $\lambda_{\text{NMAE}} \ge 0$ are hyperparameters that determine the influence of each loss component in the overall optimization process.
\section{Experimental Setup}
\label{sec:sec4}

We conducted two sets of experiments, addressing the reconstruction task from two perspectives: a 2D reconstruction problem and a 2D dynamic reconstruction problem involving spatial dimensions and time.

\subsection{2D Reconstruction}
\label{sec:subsec4.1}

In this setup, our goal was to solve \Eq{admm}. We utilized 2D U-Nets with four scales as denoisers, each featuring 32 filters in the initial scale. The optimization process involved 16 steps (T = 16). Data consistency in \Eq{admm_x} was ensured through 14 gradient descent iterations. For the sensitivity model, we employed a 2D U-Net with four scales and 32 filters for the first scale. This configuration focused on reconstructing 2D images. The input consisted of undersampled multi-coil $k$-space data from single slices or frames, and the output comprised 2D images.

\subsection{2D Dynamic Reconstruction}
\label{sec:subsec4.2}

In this configuration, we approached the reconstruction challenge dynamically, utilizing the formulation presented in \Eq{admm_3d}. Our model took as input a sequential series of time frames featuring 2D undersampled multi-coil $k$-space data. Our objective was to generate a corresponding sequential series of time-frame images as the output.  In contrast to the previous setup, we employed 3D U-Nets, incorporating four scales and 32 filters in the initial scale. However, to accommodate GPU memory constraints, we limited the optimization steps to T = 10 and conducted 8 gradient descent iterations for data consistency. Similarly to the 2D reconstruction setup, for the sensitivity model we utilized a 2D U-Net with four scales and 32 filters in the initial scale.

\subsection{Dataset}
\label{sec:subsec4.3}
We conducted our experiments using the CMRxRecon dataset \cite{wang2023cmrxrecon}, containing 4D multi-coil Cine and multi-contrast $k$-space data acquired on a 3T MRI scanner with protocols outlined in \cite{Wang2021}. The Cine MRI data included short-axis (SAX) and long-axis (LAX) views, while the multi-contrast data encompassed T1 and T2-weighted MRI data. For training, we had access to a total of 203 cine and 240 multi-contrast 4D volumes of fully-sampled $k$-spaces. The validation dataset comprised 111 cine and 118 multi-contrast 4D volumes of undersampled $k$-spaces at acceleration factors of 4, 8, and 10.

\subsection{Training and Optimization Details}
\label{sec:subsec4.4}

Our models were implemented and optimized using PyTorch \cite{paszke2017automatic}. The Deep Image Reconstruction Toolkit (DIRECT) \cite{DIRECTTOOLKIT} facilitated our pipeline tools. We employed Adam as the model parameter optimizer, with $ \epsilon=10^{-8}$ and $\left(\beta_1,\beta_2\right) = \left(0.9, 0.999 \right)$. Training was conducted on four NVIDIA A100 80GB GPUs with a batch size of 1 and 2 on each GPU, for dynamic and non-dynamic tasks, respectively. 

For both experimental setups, the loss computation used these weighting parameters: $\lambda_{\text{SSIM}} \, = \, \lambda_{1} \, = \, \lambda_{\text{HFEN}_{1}} \, = \, 1.0$, and $\lambda_{\text{NMAE}} \, = \, 3.0$.  For 2D dynamic reconstruction (\Section{subsec4.2}), we employed both versions of the SSIM loss, computed per 2D slice and across the entire sequence, and we set $\lambda_{\text{SSIM3D}} \, = \, 1.0$.

\subsection{Comparisons}
\label{sec:subsec4.5}
To evaluate our proposed methods, we compared them against two state-of-the-art  2D MRI reconstruction approaches, the Recurrent Variational Network (RecurrentVarNet) \cite{Yiasemis_2022_CVPR}, wining method in the MultiCoil MRI Reconstruction Challenge \cite{10.3389/fnins.2022.919186} and the End-to-end Variational Network (E2EVarNet), one of the top-performing solutions in the fastMRI challenge \cite{10.1007/978-3-030-59713-9_7}. Both approaches were trained using the same settings and techniques as used for our proposed methods.

\subsection{Evaluation Metrics}
\label{sec:subsec4.6}

Metrics used for evaluation were the structural similarity index measure (SSIM), the normalized mean-squared-error (NMSE), and the peak signal-to-noise ratio (PSNR). 

\section{Results}
\label{sec:sec5}

\setlength{\tabcolsep}{1.2pt}
{\renewcommand{\arraystretch}{1.5}
\begin{table}[!ht]
\centering
\setlength{\belowcaptionskip}{-15pt}
\caption{Average evaluation metrics on the validation set for each modality.}
\label{tab:results}
\resizebox{1\textwidth}{!}{%
\begin{tabular}{ccccccccccccccccccc}
\cline{1-1} \cline{3-3} \cline{5-11} \cline{13-19}
\multirow{3}{*}{\textbf{\begin{tabular}[c]{@{}c@{}}Experimental\\ Setup\end{tabular}}}              &  & \multirow{3}{*}{\textbf{\begin{tabular}[c]{@{}c@{}}Acceleration\\ Factor\end{tabular}}} &  & \multicolumn{7}{c}{\textbf{Cine}}                                                                          &  & \multicolumn{7}{c}{\textbf{Multi-Contrast}}                                                                         \\ \cline{5-11} \cline{13-19} 
                                                                                                    &  &                                                                                         &  & \multicolumn{3}{c}{\textbf{LAX}}                   &  & \multicolumn{3}{c}{\textbf{SAX}}                   &  & \multicolumn{3}{c}{\textbf{T1-weighted}}           & \textbf{} & \multicolumn{3}{c}{\textbf{T2-weighted}}           \\ \cline{5-7} \cline{9-11} \cline{13-15} \cline{17-19} 
                                                                                                    &  &                                                                                         &  & SSIM            & NMSE            & PSNR           &  & SSIM            & NMSE            & PSNR           &  & SSIM            & NMSE            & PSNR           &           & SSIM            & NMSE            & PSNR           \\ \cline{1-1} \cline{3-3} \cline{5-7} \cline{9-11} \cline{13-15} \cline{17-19} 
\multirow{3}{*}{RecurrentVarNet}                                                                    &  & 4                                                                                       &  & 0.8696          & 0.0192          & 31.07          &  & 0.9170          & 0.0118          & 34.14          &  & 0.9016          & 0.0175          & 33.21          &           & 0.8995          & 0.0125          & 31.34          \\
                                                                                                    &  & 8                                                                                       &  & 0.7871          & 0.0505          & 26.99          &  & 0.8499          & 0.0272          & 30.42          &  & 0.8360          & 0.0424          & 29.46          &           & 0.8534          & 0.0266          & 28.08          \\
                                                                                                    &  & 10                                                                                      &  & 0.7763          & 0.0592          & 26.46          &  & 0.8295          & 0.0362          & 29.24          &  & 0.8034          & 0.0601          & 27.79          &           & 0.8451          & 0.0340          & 27.04          \\ \cline{1-1} \cline{3-3} \cline{5-7} \cline{9-11} \cline{13-15} \cline{17-19} 
\multirow{3}{*}{E2EVarNet}                                                                          &  & 4                                                                                       &  & 0.9521          & 0.0048          & 37.43          &  & 0.9693          & 0.0033          & 40.67          &  & 0.9715          & 0.0038          & 41.34          &           & 0.9543          & 0.0042          & 36.64          \\
                                                                                                    &  & 8                                                                                       &  & 0.8871          & 0.0174          & 31.79          &  & 0.9262          & 0.0095          & 35.24          &  & 0.9354          & 0.0107          & 35.63          &           & 0.9261          & 0.0093          & 33.08          \\
                                                                                                    &  & 10                                                                                      &  & 0.8727          & 0.0209          & 30.79          &  & 0.9112          & 0.0126          & 33.91          &  & 0.9202          & 0.0190          & 33.21          &           & 0.9205          & 0.0114          & 32.02          \\ \cline{1-1} \cline{3-3} \cline{5-7} \cline{9-11} \cline{13-15} \cline{17-19} 
\multirow{3}{*}{\begin{tabular}[c]{@{}c@{}}2D Reconstruction\\ (2D vSHARP)\end{tabular}}            &  & 4                                                                                       &  & 0.9584          & 0.0034          & 38.74          &  & 0.9739          & 0.0025          & 41.54          &  & 0.9766          & 0.0026          & 42.16          &           & 0.9573          & 0.0038          & 36.94          \\
                                                                                                    &  & 8                                                                                       &  & 0.9072          & 0.0111          & 33.50          &  & 0.9410          & 0.0069          & 36.75          &  & 0.9521          & 0.0063          & 37.87          &           & 0.9369          & 0.0069          & 34.31          \\
                                                                                                    &  & 10                                                                                      &  & 0.8944          & 0.0138          & 32.48          &  & 0.9284          & 0.0091          & 35.50          &  & 0.9442          & 0.0092          & 36.50          &           & 0.9334          & 0.0083          & 33.57          \\ \cline{1-1} \cline{3-3} \cline{5-11} \cline{13-19} 
\multirow{3}{*}{\begin{tabular}[c]{@{}c@{}}2D Dynamic \\ Reconstruction\\ (3D vSHARP)\end{tabular}} &  & 4                                                                                       &  & \textbf{0.9658} & \textbf{0.0028} & \textbf{39.57} &  & \textbf{0.9783} & \textbf{0.0020} & \textbf{42.39} &  & \textbf{0.9814} & \textbf{0.0021} & \textbf{42.24} &           & \textbf{0.9655} & \textbf{0.0029} & \textbf{38.23} \\
                                                                                                    &  & 8                                                                                       &  & \textbf{0.9229} & \textbf{0.0087} & \textbf{34.54} &  & \textbf{0.9522} & \textbf{0.0055} & \textbf{37.81} &  & \textbf{0.9609} & \textbf{0.0055} & \textbf{38.80} &           & \textbf{0.9479} & \textbf{0.0054} & \textbf{35.47} \\
                                                                                                    &  & 10                                                                                      &  & \textbf{0.9112} & \textbf{0.0111} & \textbf{33.44} &  & \textbf{0.9407} & \textbf{0.0079} & \textbf{36.48} &  & \textbf{0.9544} & \textbf{0.0080} & \textbf{37.33} &           & \textbf{0.9460} & \textbf{0.0063} & \textbf{34.74} \\ \cline{1-1} \cline{3-3} \cline{5-11} \cline{13-19} 
\end{tabular}%
}
\end{table}
}
\setlength{\belowcaptionskip}{-20pt}
In \Figure{recons} we present sample reconstructions and in \Table{results} are presented  the reconstruction evaluation results on the validation dataset, from both of our experimental setups.  Additionally, we include results from the two methods employed for comparison: the RecurrentVarNet and the E2EVarNet. We can observe that both, 2D reconstruction and 2D dynamic reconstruction with vSHARP, yielded superior results in terms of quantitative metrics, surpassing both the RecurrentVarNet and the E2EVarNet. However, the 2D dynamic reconstruction setup outperforms the 2D reconstruction for both Cine and Multi-Contrast tasks.

Additionally, in \Table{recon-time}, we present the time required for volume reconstruction in seconds across the two experimental setups detailed in this work. From \Table{recon-time} is evident that in overall, the 2D dynamic reconstruction surpasses the 2D reconstruction in both Cine and Multi-Contrast scenarios. 

\begin{figure}[!htb]
\centering
\includegraphics[width=1\textwidth]{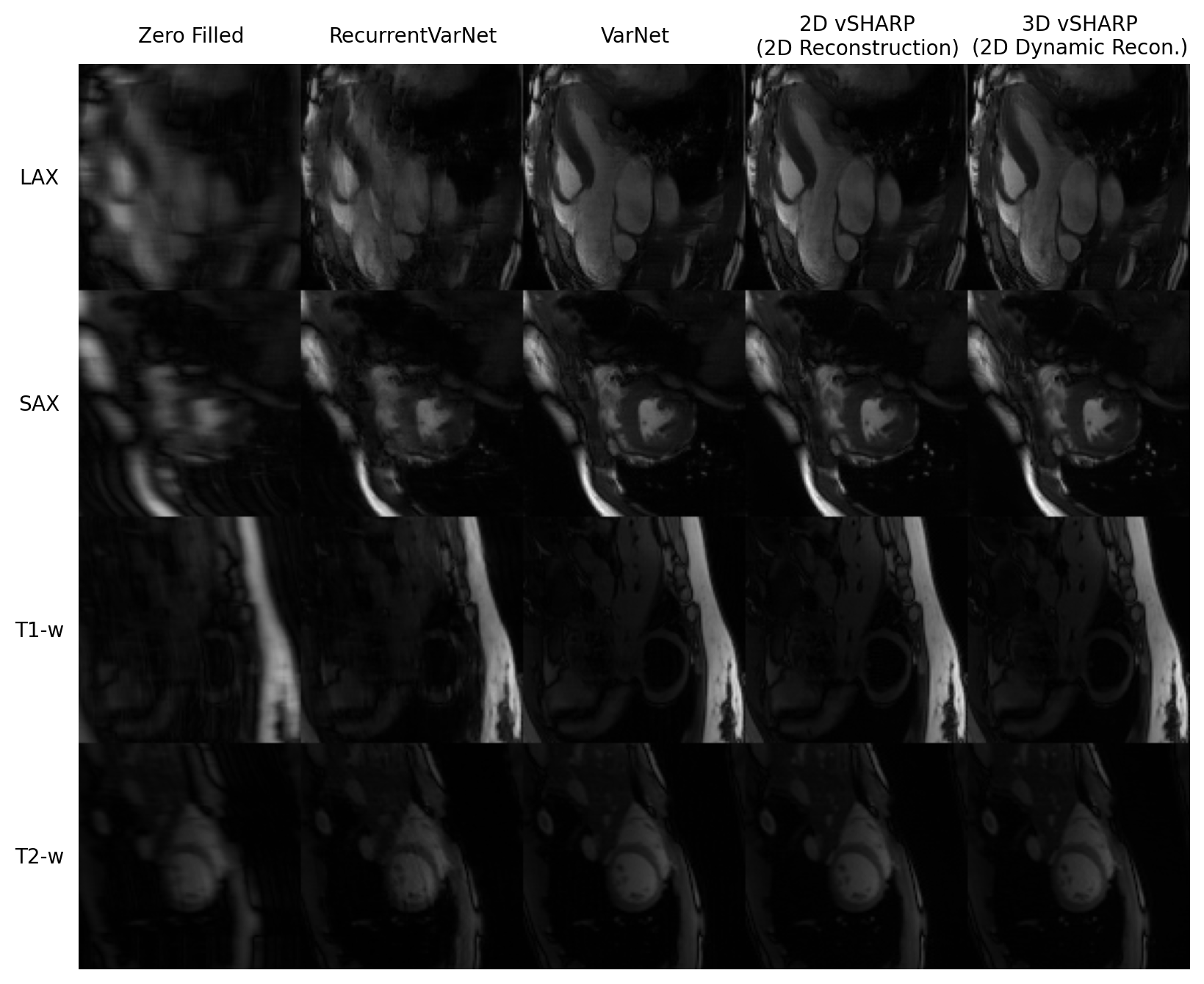}
\setlength{\belowcaptionskip}{-15pt}
\caption{Sample reconstructions from the 10$\times$ undersampled validation set.}
\label{fig:recons}
\end{figure}

\setlength{\tabcolsep}{1.2pt}
{\renewcommand{\arraystretch}{1.5}
\begin{table}[!bht]
\centering
\caption{Time for reconstruction per volume (in seconds).}
\label{tab:recon-time}
\resizebox{1\textwidth}{!}{%
\begin{tabular}{ccccccclclcccccll}
\cline{1-1} \cline{3-4} \cline{6-7} \cline{9-9} \cline{11-12} \cline{14-15}
\multirow{3}{*}{\textbf{\begin{tabular}[c]{@{}c@{}}2D \\ Reconstruction\end{tabular}}} &  & \multicolumn{2}{c}{\textbf{Cine}} & \textbf{} & \multicolumn{2}{c}{\textbf{Multi-Contrast}} &  & \multirow{3}{*}{\textbf{\begin{tabular}[c]{@{}c@{}}2D Dynamic\\ Reconstruction\end{tabular}}} &                               & \multicolumn{2}{c}{\textbf{Cine}} & \textbf{} & \multicolumn{2}{c}{\textbf{Multi-Contrast}} & \multicolumn{1}{c}{\textbf{}} & \multicolumn{1}{c}{\textbf{}} \\ \cline{3-4} \cline{6-7} \cline{11-12} \cline{14-15}
                                                                                       &  & \textbf{LAX}    & \textbf{SAX}    & \textbf{} & \textbf{T1-w}        & \textbf{T2-w}        &  &                                                                                               & \multicolumn{1}{c}{\textbf{}} & \textbf{LAX}    & \textbf{SAX}    & \textbf{} & \textbf{T1-w}        & \textbf{T2-w}        &                               &                               \\ \cline{3-4} \cline{6-7} \cline{11-12} \cline{14-15}
                                                                                       &  & 8.57            & 96.49           &           & 12.60                & 2.26                 &  &                                                                                               &                               & 3.63            & 15.71           &           & 5.46                 & 2.72                 &                               &                               \\ \cline{1-1} \cline{3-4} \cline{6-7} \cline{9-9} \cline{11-12} \cline{14-15}
\end{tabular}%
}
\end{table}
}
\section{Conclusion and Discussion}
In this work we employed the variable Splitting Half-quadratic ADMM algorithm for Reconstruction of inverse-Problems (vSHARP) network, a state-of-the-art DL-based method, to the task of reconstructing undersampled Cardiac MRI data. We adapted vSHARP under two settings, one that considers the reconstruction problem as a 2D reconstruction task, i.e., each image at a specific time frame is treated individually, and one that it considers it as a dynamic task by operating on all time frame data within a given sequence.

Upon reviewing the \Table{results}, it becomes evident that both of our proposed methods have demonstrated superior performance compared to the alternatives. In addition, as anticipated and demonstrated in other works \cite{chaoping2023}, our empirical findings confirm that 2D dynamic reconstruction outperforms the traditional 2D reconstruction. This improved performance of the 2D dynamic model can be attributed to its ability to leverage shared information across data points within the same time sequence.

Another aspect worth considering is that, in our dynamic setup, we employed all time frames per slice as input. This introduced GPU memory limitations, thereby constraining the parameter count in the reconstruction model (3D vSHARP). However, by utilizing only a subset of the time sequence data (e.g., 2-3 adjacent time frames), it would be feasible to construct a larger model.

Furthermore, \Table{recon-time} shows that the 2D dynamic reconstruction setup requires less inference time. This can be attributed to the fact that the 2D reconstruction process involves loading individual slices or time frames into memory and subsequently performing a forward pass through the model. This leads to relatively longer reconstruction times, as evidenced by the higher values for both the Cine and Multi-Contrast datasets. Conversely, in the 2D dynamic reconstruction setup, sequences of data are loaded collectively and processed in a single forward pass through the 2D dynamic model, resulting in significantly reduced reconstruction times. This observation could indeed play a pivotal role in selecting an appropriate reconstruction model for real-time clinical scenarios.

\section*{Acknowledgements}
This work was funded by an institutional grant from the Dutch Cancer Society and the Dutch Ministry of Health, Welfare and Sport.
\newpage
%
%

%
%
\bibliographystyle{IEEEtran}
\bibliography{bib}

\end{document}